\documentclass[%
 aip,
 apl,
 amsmath,amssymb,
reprint,
]{revtex4-1}

\usepackage{graphicx}
\usepackage{dcolumn}
\usepackage{bm}
\usepackage[mathlines]{lineno}

\usepackage[utf8]{inputenc}
\usepackage[T1]{fontenc}
\usepackage{mathptmx}
\usepackage[dvipsnames]{xcolor}

\emergencystretch=\maxdimen
\begin{document}


\title{Charge-pumping with finger capacitance in a custom electrostatic energy harvesting ASIC}

\author{A. Y. Zhou}
\affiliation{Department of Electrical Engineering and Computer Science, University of California, Berkeley, California, 94709, USA}
\author{M. M. Maharbiz}%
 \email{maharbiz@berkeley.edu}
\affiliation{Department of Electrical Engineering and Computer Science, University of California, Berkeley, California, 94709, USA}
 \affiliation{Department of Bioengineering, University of California, Berkeley, California, 94709, USA}
 \affiliation{Chan Zuckerberg Biohub, San Francisco, California, 94158, USA}

\begin{abstract}
We present an integrated circuit capable of scavenging energy from repetitive changes in finger touch capacitance. A finger tapping on this ASIC generates a capacitive change of approximately 770pF. These changes feed into a charge-pump circuit which stores 320pJ of energy on a 1nF storage capacitor. We present measurement results and simulations that demonstrate operation. As a proof-of-concept, we also demonstrate that the harvested energy can power a ring oscillator which outputs a series of chirps with frequencies ranging from 80Hz to 30kHz as the storage capacitor voltage charges and discharges.
\end{abstract}

\maketitle

The proliferation of Internet-of-Things (IoT) systems and shrinking consumer electronics have generated a demand for battery-less power sources for some applications. Although batteries continue to be the principal power source for most devices (of any kind), their size, cost, and maintenance requirements may limit their application scope \cite{Raghunathan2006}. As a complementary alternative, energy harvesting generators are an attractive wireless recharging option, or altogether replacement for batteries, since they scavenge energy from various ambient energy sources \cite{Mitcheson2008}. In particular, the human body is constantly doing work through various tasks \cite{Starner1996}, and harvesting fractions of this energy can be particularly useful for wearable sensors or interactive electronics.

A significant body of literature explores the conversion of kinetic energy of human body motion into useful electrical energy. The most common transduction methods include piezoelectric \cite{Li2014,Toprak2014,Liu2018}, electromagnetic \cite{Wang2017,Zhao2019}, electrostatic \cite{Zhang2016,Lu2015}, and triboelectric \cite{Ahmed2019,Wu2019}. These technologies each have strengths and weaknesses, as outlined by Zhou, et al \cite{Zhou2018}. Here, we focus on electrostatic transduction due to the ease with which these systems can be manufactured and, importantly, co-fabricated with commoditized integrated circuits (ICs). Most other energy harvesters presented in the literature use piezoelectric, magnetic, or triboelectric materials which are not available in traditional, commoditized CMOS foundry processes. As a result, delivering power from these harvesters to power management and device electronics can require additional integration steps which can be costly and inconvenient. More importantly, a paramount requirement of body energy harvesting is that the device is unobtrusive to natural motion, so additional wires and interconnects are undesirable as they can increase weight and size, while also reducing reliability and efficiency. Electrostatic harvesting requires only a variable capacitor which is readily incorporated into existing standard fabrication processes, allowing the energy harvesting module and electronics to be present on a single silicon die.

While the most powerful actions are the large movements of the limbs, which occur during activities such as walking or shaking arms \cite{Donelan2008,Qian2018,Wu2017,Li2018}, it can be inconvenient to charge devices through these motions for more sedentary persons and those in movement-restrictive environments. Users have to either walk around whenever they want to supply power, or carry a large energy storage device that collects ambient energy throughout the day. In an effort to minimize user effort, we focused in this work on leveraging natural motions people already use to interact with electronics, such as tapping, clicking, and pointing motions \cite{Wang2009}, to provide on-demand power. As is well-known for capacitive touch screen devices, finger proximity can shift the effective capacitance present between electrical nodes in a circuit \cite{Barrett2010}. This changing capacitance is the driving force behind the electrostatic energy harvester presented here.

Below, we present a custom application specific integrated circuit (ASIC) chip with built-in energy harvesting capabilities on the same silicon die. As a finger approaches the device, a change in capacitance is detected by two nodes of the ASIC. These changes feed into a rectifying charge-pump circuit which stores the energy on a capacitor. Once enough energy has been harvested and the capacitor reaches the threshold voltage of a gating switch, downstream circuits can be triggered and powered. We demonstrate switching on and powering a ring oscillator (RO) as a stand-in for generic blocks like a low power burst emitter or timing circuit.

\begin{figure}
\includegraphics{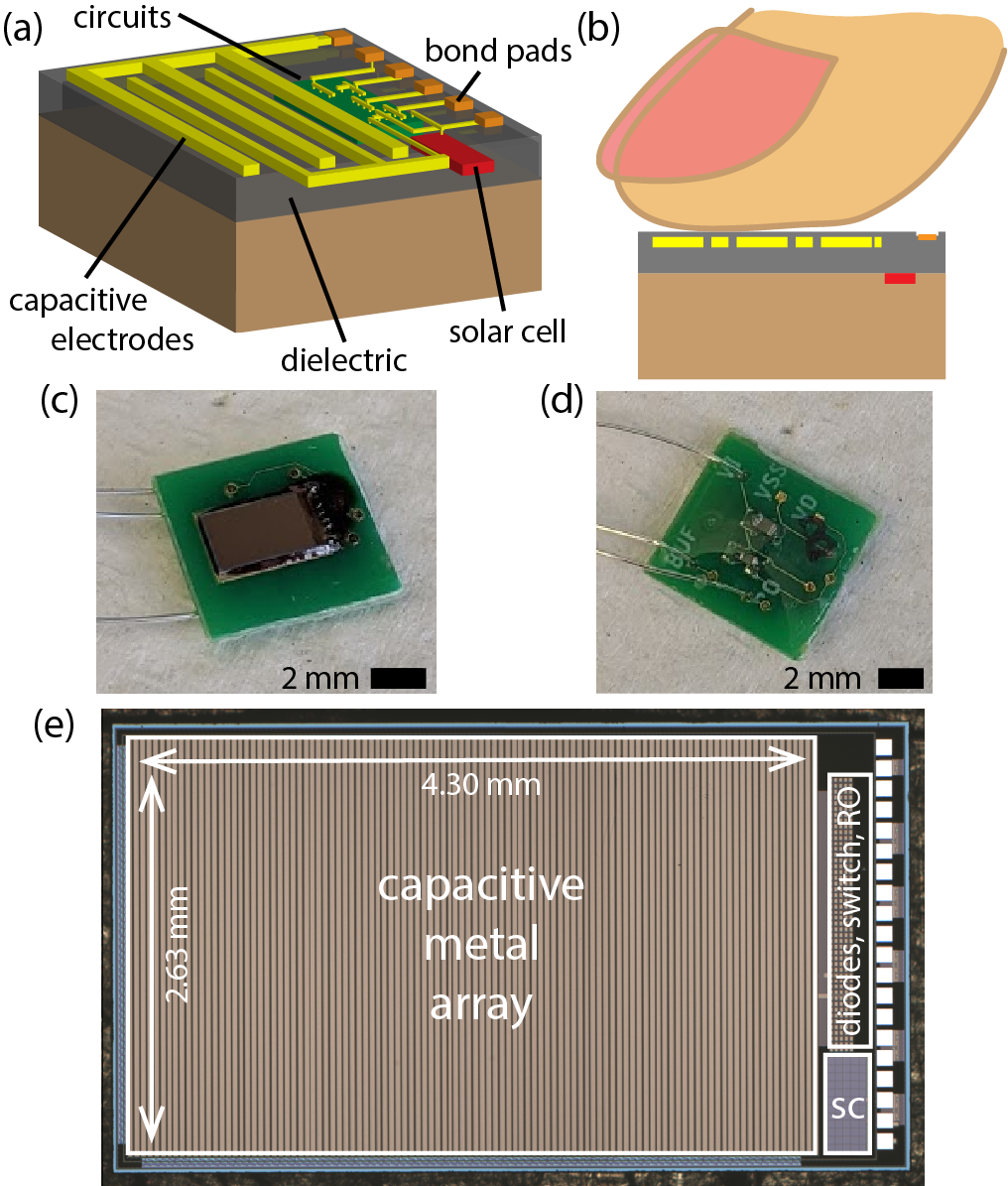}%
\caption{\label{asic} Custom ASIC with integrated electrostatic energy harvesting. (a) Cartoon schematic of the ASIC. (b) Representation of how a finger interacts with the cross-section of the ASIC during energy harvesting. (c) Fabricated chip wirebonded to a PCB for testing. (d) Backside of PCB. (e) Chip die photo.}%
\end{figure}

The ASIC, illustrated in Fig. \ref{asic}, consists of five main components: (1) top level capacitive metal traces, (2) bond pads, (3) dielectric, (4) lower level metal interconnects, and (5) active area, which are fabricated on a bulk silicon substrate in a standard 0.18$\mu$m CMOS process. The key parameters of these components are listed in Table \ref{parameters}. The top level capacitive metal traces were fabricated in metal layers three through six, while the first two metal layers were reserved for routing between the active circuit components. The process includes insulation between all the metal traces which also serves as the dielectric for the finger-sensitive capacitor. As a finger approaches the chip, it modulates the electric field between these metal traces, which create the transient changes in capacitance used to drive the energy harvester.

A well-known disadvantge of electrostatic energy harvesting is the requirement for an external power supply to initially bias the entire system. To address this challenge, we have introduced a very small integrated solar cell module that provides this initial bias voltage across the capacitors. Under indoor fluorescent lighting, which represents expected operating conditions, we characterize the solar cell with a short circuit current of approximately 47.5nA and an open circuit voltage of 0.35V. This solar cell provides charge for the initial bias in the system.

\begin{table}
\caption{\label{parameters} Identified system parameters}
\begin{tabular}{l c}
Parameter & Value \\
\hline
Total capacitive area & 11.3 mm$^2$ \\ 
Width of sense fingers & 35 $\mu$m \\ 
Width of ground fingers & 2 $\mu$m \\ 
Length of fingers & 2.6 mm \\ 
Gap between fingers & 2 $\mu$m \\
\# of fingers on ground electrode & 104 \\
\# of fingers on sense electrode & 105 \\
Circuit active area & 0.2 mm$^2$ \\ 
Solar cell area & 0.15 mm$^2$ \\
Bond pad area (single) & 0.01 mm$^2$ \\ 
Total die volume & 3.9 mm$^3$ \\ 
\end{tabular}
\end{table}

\begin{figure*}
\includegraphics[width=\textwidth]{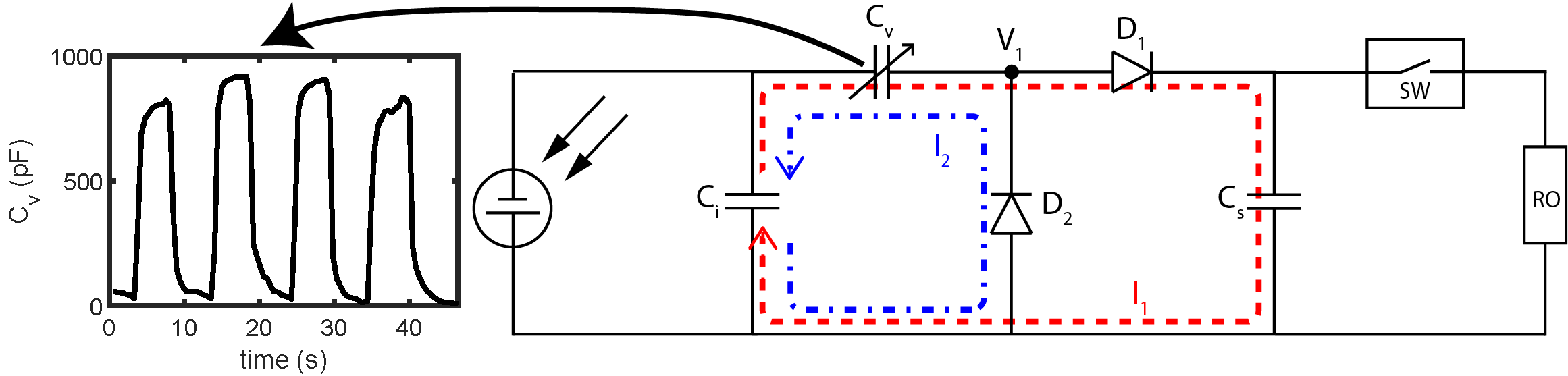}
\caption{\label{rectcircuit} Energy harvesting rectifying charge-pump circuit, with a solar cell for initial bias, delivering power to a ring oscillator (RO) through a diode-connected MOSFET switch (SW). Cv represents the variable capacitor in the system, which changes as a finger approaches the top metal traces on the ASIC. The left graph shows capacitance measured by an LCR meter over four slow finger taps.}%
\end{figure*}

During operation, a finger taps aperiodically on the ASIC and influences the effective capacitance seen at the top level metal traces. These capacitance changes are measured with an LCR meter (Keysight E4980a) and shown in Fig. \ref{rectcircuit} (left). The nominal capacitance of the traces alone is $\sim$103pF, and each finger tap results in an increase in capacitance to around 825-920pF. As with any human interactive device, we see a wide range of resulting capacitances since these change with finger placement, timing, and operating conditions, such as dirt and sweat. For designing and analyzing our proof-of-concept system, the average value of Cmax=870pF was used in all simulations. This change in capacitance feeds into the rectifying charge-pump circuit presented in Fig. \ref{rectcircuit} (right). During a charging cycle, the voltage at V1 increases; this turns diode D1 on and pumps charge onto the storage capacitor Cs through the current pathway I1. On the opposing edge of the variable capacitor, V1 decreases; this restores the initial capacitor, Ci, and variable capacitor, Cv, to their original states via charge transfer through diode D2 and current pathway I2. To ensure minimum leakage in the rectifying circuit, both diodes were isolated with high-voltage (deep) n-wells and substrate grounding. In this version, both the initial and storage capacitor are incorporated as discrete, surface-mount components for flexibility of testing various values. For a fully integrated solution, smaller capacitors can be fabricated on chip.

\begin{figure}
\includegraphics[width=\columnwidth]{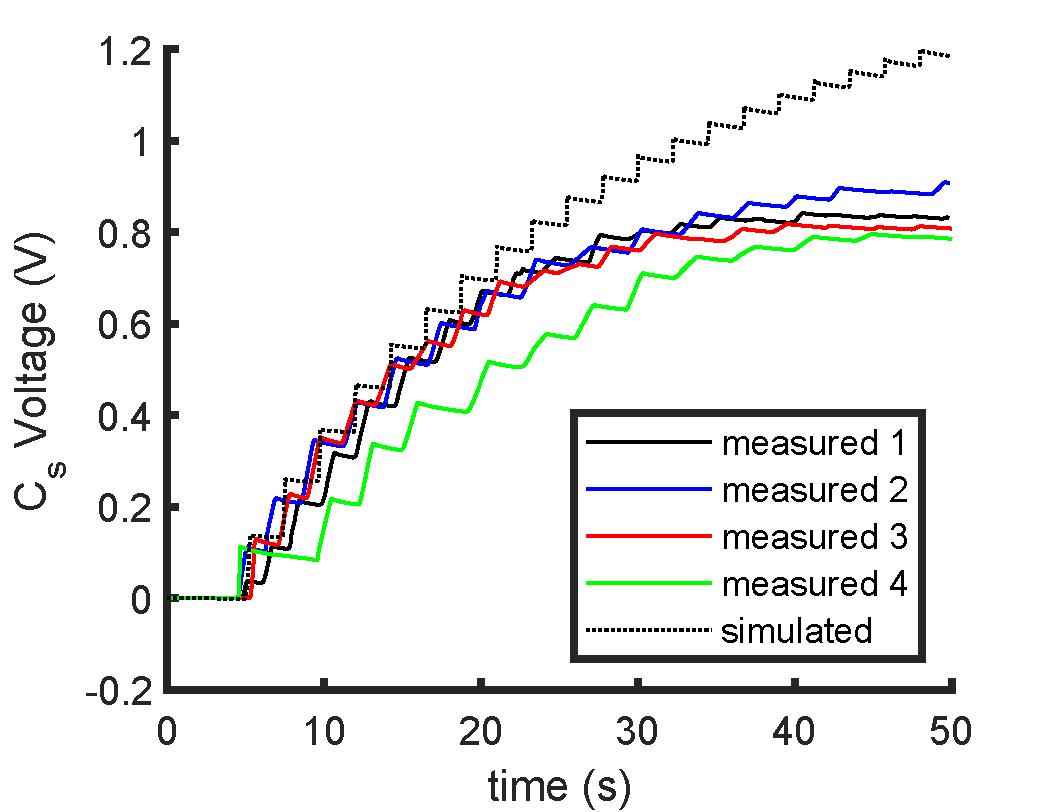}%
\caption{\label{Cs} Four examples of the measured voltage (solid) and simulated voltage (dotted) on the storage capacitor during energy harvesting. Each tap pumps charge onto Cs creating a staircase-like curve.}%
\end{figure}

To demonstrate the energy harvester powering an on-chip circuit application, we implemented a diode-connected MOSFET switch with a trigger voltage of $\sim$650mV. Once the charge on the storage capacitor is sufficient and the capacitor voltage exceeds the switching threshold, the 17-stage RO turns on and outputs a series of short chirps. Circuit performance was simulated with Cadence Spectre transient analysis and its results are presented along with experimental data in Fig. \ref{Cs}.

To assemble the device for testing purposes, the ASIC was wirebonded to a custom printed circuit board (PCB) with test leads out to a sourcemeter (Keithley 2400) or digital oscilloscope (National Instruments USB-5133). The wirebonds were mechanically protected and electrically isolated with a low viscosity epoxy (EPO-TEK 353ND), which wicks up the length of the wirebond without significant spreading on the surface of the chip. Future iterations of this device can use bumping assembly methods to avoid potential wirebond breakage and reduce area requirements. In any assembly process, specific care should be taken to avoid coating the ASIC with unwanted materials, which increases the distance between the finger and metal capacitive traces, reducing the effective change in capacitance. In our work, all exposed metal on the component side of the PCB was also insulated with UV curable epoxy (EPO-TEK OG116-31) to prevent shorting through the skin during finger contact.

We first measured the energy harvested over a series of finger taps by tracking the voltage across a storage capacitor (Cs) with a voltmeter. During these experiments, the capacitor (Cs=1nF) was not connected to any application loads. We tapped the chip at a rate of $\sim$25 taps per min and saw a staircase-like increase of voltage on Cs (Fig. \ref{Cs}). As expected with a human interactive device, we saw multiple energy harvesting pathways for each experimental measurement due to variations in tapping frequencies, finger placement, and finger conditions. Overall, we saw good agreement with the simulations.  Initial taps show a voltage increase on the storage capacitor of $\sim$0.12V. Using the relationship $Q = CV$, this means $\sim$120pC of charge was pumped onto the storage capacitor during a single tap. Since we know the energy of a storage capacitor is $E = \frac{1}{2}CV^2$, we can solve for $\sim$7.2pJ of energy transferred to Cs. The energy transfer decreased upon successive taps because as the charge on Cs increased, the voltage differential across diode D1 decreased, which decreased the current and charge transferred on each subsequent tap. After around 50 seconds of tapping at this frequency, Cs reached a total stored energy of 320pJ at 0.8V. 

Given this data, we calculated approximate power and power density for this device. Using a leisurely tapping frequency of 25 taps per min, or 2.4 seconds per tap, the device delivered $\Delta P_{max}=\frac{\Delta E}{\Delta t}=\frac{7.2pJ}{2.4s}=3pW$ of power. We can also solve for power density using the chip die volume of 3.9mm$^3$, which results in a power density of 769pW/cm$^3$. Another way to increase power delivered by the device is by increasing the variable capacitor area; however, because the finger capacitance is limited by the total area available on the ASIC, integrating the touch capacitor onto the chip limits the realizable change in capacitance. If some assumptions are made about the finger movement, we can also approximate the energy conversion efficiency of this system. If we estimate the tip of a finger to be a 1cm diameter sphere of water, the weight of this sphere is $\sim$0.5g. Assuming a simple linear motion, $W=F\times d$ and $F=ma$, we find that the work exerted is $\sim$50$\mu$J of energy for a distance traveled of 1cm. This would mean an energy conversion efficiency of the system defined as $\textit{eff} = \frac{E_{C_s}}{E_{tap}}\times 100\%=\frac{7.2pJ}{50\mu J}\times 100\% = 1.4\times 10^{-5}\%$. We expect this efficiency to be very small because even though the finger travels multiple millimeters through space in each tap, only the last couple tens to hundreds of microns do work and effect charge movement in the capacitive electrodes; therefore, significant amounts of kinetic energy are not converted.

\begin{figure}
\includegraphics{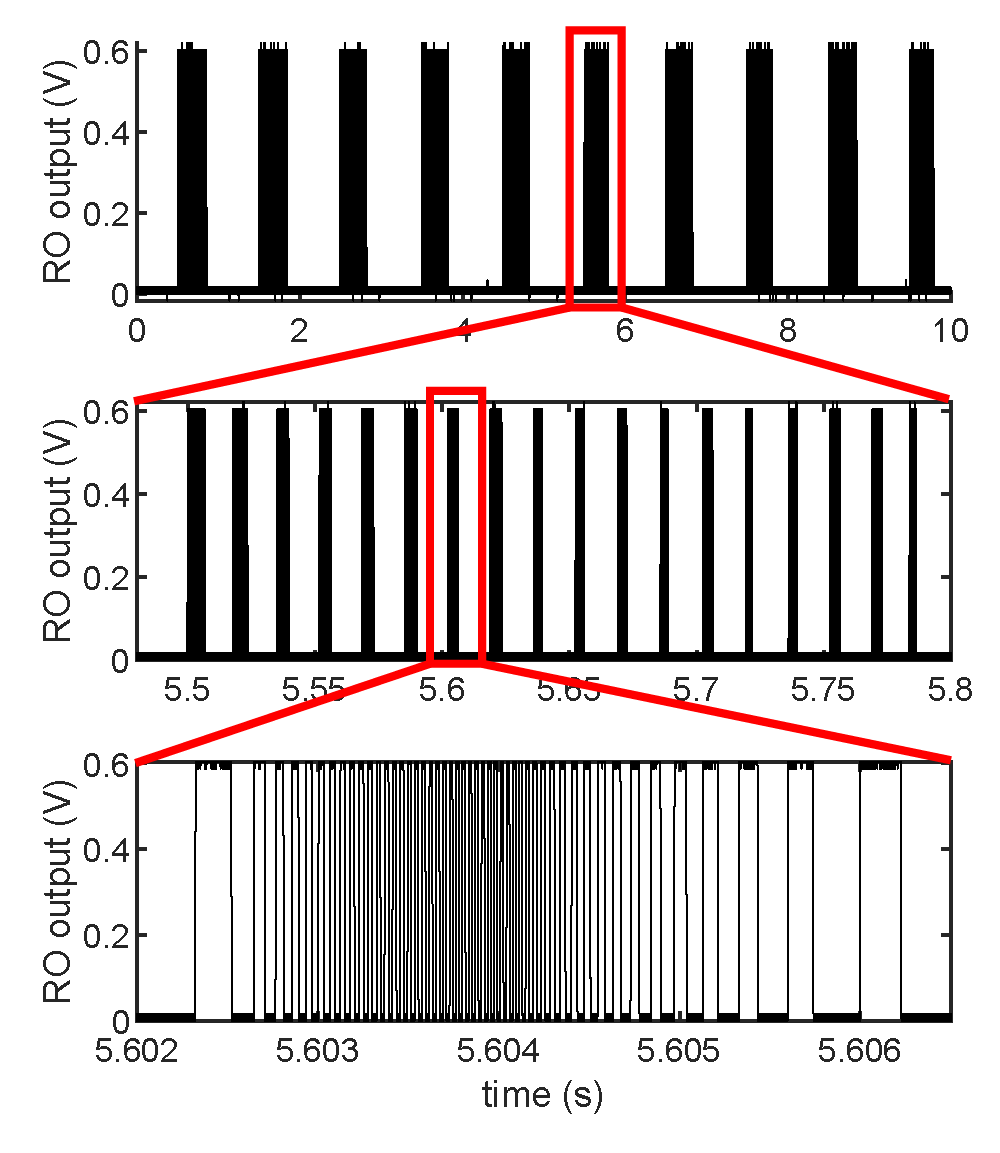}%
\caption{\label{RO} The RO output signal from being powered by charge-pumping with finger capacitance. Ten finger taps produces an activation of the RO (top). Each tap consists of a series of chirps (middle). Each chirp varies in frequency with modulating Cs voltage.}%
\end{figure}

After validating operation and measuring available power, we tested the ability of the energy harvester to power a simple on-chip ring oscillator (RO). We directly connected the storage capacitor (Cs = 10pF) to a diode-connected MOSFET, which acted as a switch and drew very little current at low voltages. Once enough charge was pumped onto Cs, so that the voltage exceeded the switch trigger level of $\sim$650mV, the switch closed and powered the RO. The oscillator ran and drained energy from Cs until it switched off (when its voltage dropped below threshold). When the RO first turned on, we saw an increase in frequency as the voltage on Cs increased. As cycles continued to get shorter, the RO pulled more charge from the capacitor and therefore the voltage on Cs fell, leading to a decrease in frequency. This push and pull between harvested charge from finger capacitance and current draw from the RO resulted in each finger tap creating a series of chirps. In order to record these chirps, we used an externally-powered buffer to drive the signal into an oscilloscope. The results of ten finger taps are shown in Fig. \ref{RO} with subplots showing a zoomed-in view of a single tap, followed by a single chirp. The RO output frequency ranged from 80Hz to 30kHz. Knowing the power consumption of the RO from simulations, this measured output implies the load in the system consumed between 37pW (at 80Hz) to 9.4nW (at 30kHz) of power. Since RO frequency was proportional to voltage, we calibrated the RO and calculated that these frequencies correspond to harvested Cs voltages of 625-890mV. We saw that this trigger voltage was lower than our simulated 650mV, which was likely due to a combination of process variation during chip fabrication and power draw from the electronic switch. A mechanical switch with zero open-circuit power consumption could replace this transistor to maximize the power delivered to the application. This system demonstrated a proof-of-concept for automatic activation of an electronic circuit upon integrated electrostatic energy harvesting.

In conclusion, we presented an integrated system combining electrostatic energy harvesting with electronics in a single standard CMOS fabrication process. This device harvested kinetic energy from a finger tap to power a ring oscillator. As with any electrostatic energy scavenger, harvesting is limited by resistive and capacitive parasitics. These losses can be reduced with use of more advanced fabrication technologies, such as dielectric trench isolation or native silicon isolation which were not available in our technology. Since harvested energy increases with area, this system is also constrained by area limitations of the CMOS fabrication process. However, the capacitive electrodes can be easily moved off-chip to various conductive mediums such as PCB traces \cite{Zhou2017}, conductive fabric, or conductive paints on paper. This allows for larger capacitive areas which results in more energy harvested for applications which may require it. We envision embedding a series of these energy scavenging motes into everyday objects, such as books, textiles, or other consumer goods, to create smart, interactive devices.

\begin{acknowledgments}
The authors are grateful to the TSMC University Shuttle Program, which provided chip fabrication through the Berkeley Sensor and Actuator Center. The authors would also like to acknowledge M.M. is a Chan Zuckerberg Biohub investigator.
\end{acknowledgments}

\bibliography{SkinChipfull}

\end{document}